\documentclass[aps,prl,twocolumn, showpacs,nofootinbib,preprintnumbers,groupedaddress]{revtex4}
\usepackage{amssymb}
\usepackage{setspace}
\usepackage{amsmath}
\usepackage{latexsym}
\usepackage{graphicx}
\usepackage{graphics}

\begin{document}

\title{Surface alignment and anchoring transitions in nematic lyotropic
chromonic liquid crystal.}
\author{V. G. Nazarenko$^{1}$}
\author{O. P. Boiko$^{1,2}$}
\author{H.-S. Park$^{2}$}
\author{O. M. Brodyn$^{1}$}
\author{M. M. Omelchenko$^{2}$}
\author{L. Tortora$^{2}$}
\author{Yu. A. Nastishin$^{2,3}$}
\author{O. D. Lavrentovich$^{2}$}
\email{olavrent@kent.edu}
\affiliation{$^{1}$ Institute of Physics, prospect Nauky 46, Kiev-39, 03039, Ukraine\\
$^{2}$ Liquid Crystal Institute and Chemical Physics Interdisciplinary
Program, Kent State University, Kent, OH 44242 \\
$^{3}$Institute of Physical Optics, 23 Dragomanov str., Lviv, 79005, Ukraine}
\date{\today }

\begin{abstract}
The surface alignment of lyotropic chromonic liquid crystals (LCLCs) can be
not only planar (tangential) but also homeotropic, with self-assembled
aggregates perpendicular to the substrate, as demonstrated by mapping
optical retardation and by three-dimensional imaging of the director field.
With time, the homeotropic nematic undergoes a transition into a tangential
state. The anchoring transition is discontinuous and can be described by a
double-well anchoring potential with two minima corresponding to tangential
and homeotropic orientation.
\end{abstract}

\maketitle

\bigskip \newpage

Spatial bounding of a liquid crystal (LC) lifts degeneracy of molecular
orientation specified by the director $\widehat{\mathbf{n}}$ and sets an
"easy axis" $\widehat{\mathbf{n}}_{0}$ at the surface. Deviation of $%
\widehat{\mathbf{n}}$ from $\widehat{\mathbf{n}}_{0}$ requires some work
thus establishing a phenomenon of "surface anchoring" that has been explored
extensively for thermotropic LCs \cite%
{Horn,Cognard,YangRosenblatt,Volovik,Barbero,YokoyamaVan,Sluckin,Bechhoefer,Book,Patel,Sheng}%
. For lyotropic LCs, such as water solutions of polyelectrolytes,
surfactants, dyes, etc., the studies of anchoring are scarce. The view is
that the surface alignment of lyotropic LCs is determined by an excluded
volume effect, which favors the longest dimension of building units to be
parallel to a substrate \cite{Meyer,Poniewierski,Sharlow,Poulin}. We study
surface phenomena in nematic lyotropic chromonic LCs (LCLCs), a distinct
class of self-assembled LCs formed by water solutions of plank-like
molecules with polyaromatic cores and ionic peripheral groups \cite{Lydon}.
Reversible chromonic assembly and mesomorphism are displayed broadly by
dyes, drugs and nucleotides \cite{Lydon}. In water, the LCLC molecules stack
face-to-face, forming elongated aggregates. The aggregates are not fixed by
covalent bonds, being polydisperse with an average length $l\varpropto \sqrt{%
\phi }\ln \left( E/k_{B}T\right) $ that depends on temperature $T$, volume
fraction $\phi $, and stacking energy $E\sim \left( 4-10\right) k_{B}T$ \cite%
{Dickinson}. We demonstrate that in LCLCs, $\widehat{\mathbf{n}}_{0}$ can be
either parallel to a substrate (planar or tangential alignment, denoted "P")
or perpendicular (homeotropic, or H alignment), with discontinuous
transitions between the two, thus suggesting that both entropy and
anisotropic molecular interactions control the surface phenomena.

We study disodium cromoglycate (DSCG) \cite{Lydon}, $C_{23}H_{14}O_{11}Na_{2}
$ (Spectrum Inc, purity $98{\%}$), dissolved in water at 15 wt \% (mixture
A) and 12.5wt\% doped with 1.5wt\% of Na$_{2}$SO$_{4}$ (mixture\ B). The H
alignment was achieved by treating glass plates with 1\% water solution of
N,N-dimethyl-N-octadecyl-3-aminopropyl trimethoxysilyl chloride (DMOAP) \cite%
{Cognard}. The two plates are separated by Mylar strips; the cell thickness $%
d$ was measured by light interference technique. The cells were filled at $%
T_{NI}+10$ K, sealed with a UV-cured Norland epoxy glue, and cooled down to $%
T=298$ K with a rate $5$ K$/\min $ in a thermal stage HS-1 (Instec, Inc.).
We used an LC PolScope for in-plane mapping of optical retardation $R\left(
x,y\right) ={\int\limits_{0}^{d}{\left\vert n_{o}-{n_{eff}}\right\vert dz}}$%
, where ${n_{eff}}=\left( {n_{o}^{-2}\cos ^{2}\theta }+{n_{e}^{-2}\sin
^{2}\theta }\right) ^{-1/2}$, $\ \theta $ is the angle between $\widehat{%
\mathbf{n}}$ and the normal $\widehat{\mathbf{z}}$ to the cell, $n_{o}$ and $%
n_{e}$ are the ordinary and extraordinary refractive indices, respectively.
\ At 546 nm and $T=298$ K, we determined $n_{o}=1.37\pm 0.01$ and $\Delta
n_{A}=n_{o}-n_{e}=0.019\pm 0.002$ for A and $\Delta n_{B}=0.015\pm 0.002$
for B \cite{Nastishin}. To image $\widehat{\mathbf{n}}$ $\left( x,y,z\right)
$, we used fluorescence confocal polarizing microscopy (FCPM), by doping the
LCLC with 0.003 wt.\% of fluorescent acridine orange (AO, Sigma-Aldrich) and
probing it with a focused laser beam \cite{Pramana}. The fluorescence
intensity depends on the angle between $\widehat{\mathbf{n}}$ and
polarization $\mathbf{P}$ of light, being maximum for $\mathbf{P}$ $\perp $ $%
\widehat{\mathbf{n}}$ and minimum for $\mathbf{P}$ $\parallel \widehat{%
\mathbf{n}}$, suggesting that AO intercalates between the DSCG molecules.

\begin{figure}[b]
\centering
\includegraphics[width=0.48 \textwidth]{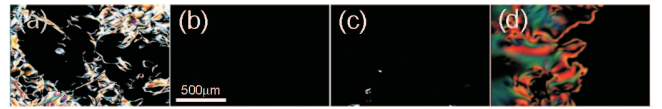}
\caption{(Color online) Textural evolution viewed between crossed
polarizers; mixture A, $d=$50 $\protect\mu $m, $T=298$ K. Dark nuclei of the
H state appear at $\protect\tau \approx $10 min after the isotropic-nematic
transition (a), H orientation at $\protect\tau $=25 min (b); appearance, $%
\protect\tau $=670 min (c) and expansion, $\protect\tau $=810 min (d) of
bright P regions. }
\label{Textural_evolution}
\end{figure}

\begin{figure}[tbp]
\centering \includegraphics[width=0.48 \textwidth]{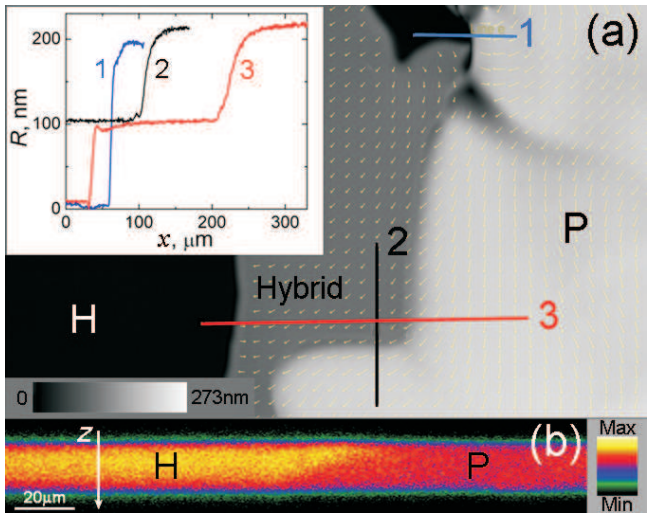}
\caption{(Color online) H-P transition in the mixture A viewed as (a) a grey
scale map of$\ R$ in $\left( x,y\right) $ plane of \ the cell ($d\approx $11
$\protect\mu $m) and as a variation of $R$ along the lines 1,2, and 3
(inset); b) an FCPM vertical cross section of a cell \ with a tilted
boundary.}
\label{Retardation_map}
\end{figure}

\indent The initial unaligned texture coarsens and then shows dark expanding
nuclei of the H state that fill the entire cell, Fig.\ref{Textural_evolution}%
. The H alignment is stable as verified by applying a strong magnetic field,
up to 7 kG, to tilt $\widehat{\mathbf{n}}$. Once the field is switched off,
the H orientation is restored. After a certain time $\tau _{H}$ $\approx $%
10-20 hours, the LCLC undergoes an H-P transition through nucleation and
expansion of birefringent domains, Fig.\ref{Textural_evolution}c,d. These
appear not only at the periphery but also in the middle of samples, Fig.\ref%
{Textural_evolution}c. Two similar cells, one left under normal conditions
and another one immersed in a mineral oil, demonstrated similar evolution.
Therefore, a possible slow drying is not a major contributor to the effect,
although the dynamics of aggregate assembly most certainly is. The H-P
transition might be direct, with $R$ abruptly changing from 0 to $R_{\max
}=\Delta n_{A}d$, line 1, or indirect, with an intermediate step $R\approx
\Delta n_{A}d/2$, lines 2, 3 in Fig.\ref{Retardation_map}a. The tilt $\beta
=\partial R/\partial x$ at the states' boundaries varies broadly, from $\sim
$100 nm/$\mu $m, to $\sim $1 nm/$\mu $m. FCPM of the vertical cross sections
shows that the boundaries represent sharp walls that are either vertical
(large $\beta $) or tilted (small $\beta $). For example, Fig.\ref%
{Retardation_map}b shows a tilted ($\sim 30^{o}$) boundary separating an H
layer with $\widehat{\mathbf{n}}||\widehat{\mathbf{z}}$ and a P layer with $%
\widehat{\mathbf{n}}\bot \widehat{\mathbf{z}}$; at either of the two H
plates, the transition from $\widehat{\mathbf{n}}||\widehat{\mathbf{z}}$ to $%
\widehat{\mathbf{n}}\bot \widehat{\mathbf{z}}$ is abrupt.

\begin{figure}[tbp]
\centering \includegraphics[width=0.48 \textwidth]{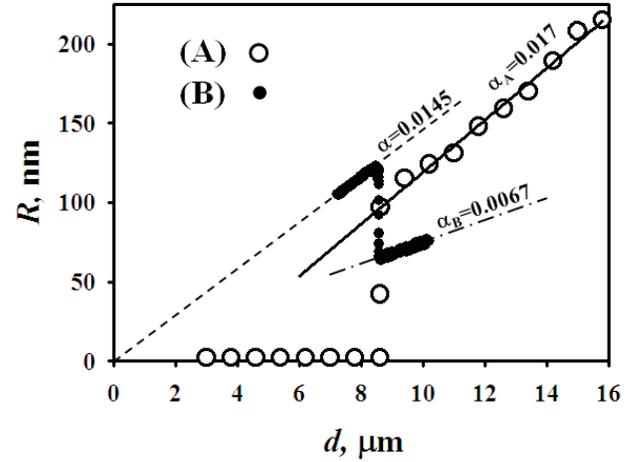}
\caption{$R(d)$ for mixtures A and B in the hybrid aligned wedge cells.}
\label{Phase_retardation_15}
\end{figure}

\begin{figure}[b]
\centering \includegraphics[width=0.48 \textwidth]{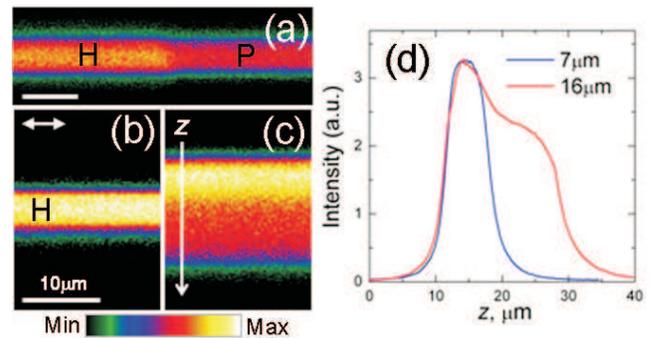}
\caption{(Color online) Vertical FCPM views of hybrid aligned wedges with
mixture A; (a) vertical boundary between H and P states, $d=10$ $\protect\mu
$m; (b) homeotropic thin, $d=7$ $\protect\mu $m and (c) hybrid thick, $d=16$
$\protect\mu $m, parts of the same wedge and\ comparison of their
fluorescent profiles (d). The double arrow shows light polarization $\mathbf{%
P}$.}
\label{FCPM}
\end{figure}

\indent To quantify the surface properties further, we use hybrid aligned
wedge cells \cite{Barbero}, assembled from two\textit{\ different} plates,
an H plate with DMOAP and a P plate with buffed polyimide SE-7511 (Nissan).
The dihedral angle is small, $<0.1^{o}$. The cells show a critical thickness
$d_{c}$ at which $R$ changes abruptly, Fig.\ref{Phase_retardation_15}; $%
d_{c} $ varies in the range 5-10 $\mu $m from sample to sample. At $d<d_{c}$%
, $\widehat{\mathbf{n}}$ is uniform, either in the H state, if the
experiment is performed at the beginning of $\tau _{H}$ for the mixture A,
or in the P state for the mixture B. At $d>d_{c}$, $\widehat{\mathbf{n}}$ is
distorted, as evidenced by the tilt $\alpha =\partial R/\partial d$ that is
neither 0 nor $\Delta n$, Fig.\ref{Phase_retardation_15}. The behavior of $R$
$\left( d>d_{c}\right) $ is complex and depends on the tilt of walls
separating the states $\widehat{\mathbf{n}}=const$ and $\widehat{\mathbf{n}}%
\neq const$. The walls can be vertical, Fig.\ref{FCPM}a, tilted (as in Fig.%
\ref{Retardation_map}b), or practically horizontal, Fig.\ref{FCPM}c. We
compare the cross sections of thin and thick parts of the same wedge, Fig.%
\ref{FCPM}b,c,d. The thin part is in a uniform H state, with fluorescence
equally strong for any in-plane orientation of $\mathbf{P}$, Fig.\ref{FCPM}%
b. In the thick part, Fig.\ref{FCPM}c, the top 1/3 is occupied with an H
layer, as evidenced in Fig.\ref{FCPM}d by an overlap of the fluorescence
profiles. \ In the bottom 2/3, $\widehat{\mathbf{n}}$ is close to planar, as
the fluorescence is weak, Fig.\ref{FCPM}c,d. \

\indent The experiments suggest that in the studied LCLC, $\widehat{\mathbf{n%
}}_{0}$ can be either tangential (planar) or homeotropic. The H alignment is
stable only within a finite period of time $\tau _{H}$. The anchoring
transitions are strongly discontinuous. The findings are unusual, as
lyotropic LCs are notoriously hard to align and when they do align, there
are no anchoring transitions. The H-P transitions occur in thermotropic LCs,
but they are continuous there \cite{Volovik,Patel}. Discontinuous
transitions were reported for \textit{in-plane} realignment at anisotropic
crystalline substrates \cite{Horn,Bechhoefer} and for patterned plates with
spatially varying easy axis \cite{Sheng}. As shown by Sluckin and
Poniewierski \cite{Sluckin}, the simplest potential leading to the
first-order H-P transitions in semi-infinite samples is $f_{s}=W_{2}(%
\widehat{\mathbf{n}}\cdot \widehat{\mathbf{z}})^{2}+W_{4}(\widehat{\mathbf{n}%
}\cdot \widehat{\mathbf{z}})^{4}$. This form also describes well the
reorientation of thermotropic LCs by external fields \cite%
{YangRosenblatt,YokoyamaVan}. For the H plate with $\widehat{\mathbf{n}}%
_{0}||\widehat{\mathbf{z}}$, $f_{s}$ can be cast as $W_{2H}\sin ^{2}\theta
_{H}+W_{4H}\sin ^{4}\theta _{H}$ with the anchoring coefficients $W_{2H}>0$
and $W_{4H}>-W_{2H}$. The change from $W_{4H}>-W_{2H}$ to $W_{4H}<-W_{2H}$
corresponds to the first-order H-P transition that has not been observed so
far experimentally \cite{Book}. Sluckin and Poniewierski \cite{Sluckin}
expressed $W_{2H}$ and $W_{4H}$ through the temperature-dependent scalar
order parameter $S(T)$ so that the transitions were temperature-driven. In
our case, the changes in $W_{2H}$ and $W_{4H}$ can be related also to the
dynamics of self-assembly. Since the H state is observed after the samples
are cooled down from the isotropic phase, it should be accompanied by the
growth of aggregates, as $l\varpropto \sqrt{\phi }\ln \left( E/k_{B}T\right)
$. Short and long aggregates might align differently at the substrates, say,
normally and tangentially, the entropy effect being one of the reasons.

For cells of a finite $d$, such as the hybrid aligned wedges, the free
energy per unit area should include both anchoring and elastic terms:
\begin{widetext}
\begin{equation}
f=\frac{1}{2}\int_{0}^{d}K(\theta )\left( \frac{d\theta }{dz}\right)
^{2}dz+W_{2H}\sin ^{2}\theta _{H}+W_{4H}\sin ^{4}\theta _{H}+W_{2P}\cos
^{2}\theta _{P}+W_{4P}\cos ^{4}\theta _{P},
\label{free_energy}
\end{equation}
\end{widetext}\noindent where $K\left( {\theta }\right) ={K_{1}\sin
^{2}\theta +K_{3}\cos ^{2}\theta }$, ${K_{1}}$ and ${K_{3}}$ are the splay
and bend elastic constants; $W_{2P}>0$ and $W_{4P}>-$ $W_{2P}$, to guarantee
$\widehat{\mathbf{n}}_{0}\bot \widehat{\mathbf{z}}$ at the P plate. For $%
d\rightarrow \infty $, the conflicting boundary conditions are satisfied by
reorienting $\widehat{\mathbf{n}}$ by $\pi /2$. As $d$ decreases, the
elastic torque $\propto \left( \theta _{P}-\theta _{H}\right) /d$,
determined by the actual polar angles $\theta _{P}$ and $\theta _{H}$ at the
plates, becomes stronger, forcing $\theta _{P}$ and $\theta _{H}$ to deviate
from their "easy" values $\pi /2$ and $0$. At some $d_{c}$, the plate with
weaker anchoring might give up, allowing $\widehat{\mathbf{n}}$ to be
uniform. \ For analytical treatment, we assume ${K_{1}}={K_{3}=K}$ and fix $%
\theta $ at the stronger anchored plate, as supported by FCPM, Fig.\ref%
{Retardation_map}b, \ref{FCPM}.

\indent For the case A in Fig.\ref{Phase_retardation_15}, $\theta _{H}=0$,
thus $f=f_{P}\left( \theta _{P}\right) =K\theta _{P}^{2}/(2d)+W_{2P}\cos
^{2}\theta _{P}+W_{4P}\cos ^{4}\theta _{P}$. The values of $\theta _{P}$
that minimize $f_{P}$ are found from the conditions $\partial f_{P}/\partial
\theta _{P}=0$ and $\partial ^{2}f_{P}/\partial \theta _{P}^{2}>0$. When $%
d\rightarrow \infty $ and $W_{4P}<-W_{2P}/2$, $f_{P}$ has an absolute
minimum at $\theta _{P}=\pi /2$ and a local one at $\theta _{P}=0$. The two
are separated by a barrier at $\theta _{Pb}=\arccos \sqrt{-\frac{W_{2P}}{%
2W_{4P}}}$. For a finite $d>>K/W_{2P}$, $f_{P}$ with $-1<W_{4P}/W_{2P}<-1/2$
preserves its double-well features. In particular, $\theta _{P}=0$ is still
a local minimum. The coordinate $\theta _{P,\min }$ of the absolute minimum,
however, becomes smaller than $\pi /2$, because of the elastic torque $%
\propto \theta _{P}/d$. We evaluate $f_{P}$ near $\theta _{P}\approx $ $\pi
/2$ to find $\theta _{P,\min }\approx \frac{\pi }{2}\left( 1-\frac{K}{%
2dW_{2P}+K}\right) $. The difference $\Delta f_{P}=f_{P}\left( \theta
_{P\min }\right) -f_{P}\left( 0\right) \ $vanishes at $d_{cP}\approx \frac{%
\pi ^{2}K}{8\left( W_{2P}+W_{4P}\right) }$. For $d<d_{cP}$, the uniform H
state is stable, while for $d>d_{cP}$, the hybrid state has the lowest
energy. The transition is discontinuous, with a big jump $\Delta \theta
_{P}\approx \frac{\pi ^{3}}{8\left( 1+W_{4P}/W_{2P}\right) +2\pi ^{2}}$ in
the range $1.31\leq \Delta \theta _{P}\leq $ $1.57$ that corresponds to the
limits $-1<W_{4P}/W_{2P}<-1/2$. Similarly for the case B, $\theta _{P}=\pi
/2 $ and $f_{H}\left( \theta _{H}\right) =K\left( \pi /2-\theta _{H}\right)
^{2}/(2d)+W_{2H}\sin ^{2}\theta _{H}+W_{4H}\sin ^{4}\theta _{H}$. At $d$
smaller than $d_{cH}\approx \frac{\pi ^{2}K}{8\left( W_{2H}+W_{4H}\right) }$%
, the uniform P state $\theta \left( z\right) =\pi /2$ is stable, and at $%
d>d_{cH}$, the hybrid state with $\theta _{H\min }=\frac{\pi }{2}\frac{K}{%
2dW_{2H}+K}<<1$ is stable. The transition is discontinuous, with the jump $%
\Delta \theta _{H}\approx \frac{\pi ^{3}}{8\left( 1+W_{4H}/W_{2H}\right)
+2\pi ^{2}}$ in the range (1.31-1.57). \ The qualitative features of this
analytical description remain intact when the full form of $f$ in\ Eq.(\ref%
{free_energy}) is analyzed numerically.

\indent The model above explains why the thin parts of wedge cells are
uniform and why the director orientation at the surfaces is close to 0 and $%
\pi /2$. The thickness $d_{c}\sim K/(W_{2}+W_{4})$ has the meaning of
surface extrapolation length, also called the de Gennes-Kleman length. The
value $d_{c}\sim 10$ $\mu $m is of the same order as the one extracted from
the experiments on elastic distortions around colloidal inclusions in LCLCs
\cite{Shiyan}. With $K\sim 10$ pN \cite{Nastishin_ELC}, one estimates $%
W_{2}+W_{4}\sim 10^{-6}$ J/m$^{2}$, which is comparable to the anchoring
coefficients found in thermotropic LCs in the regime of\ "weak" anchoring
\cite{Cognard}. The small $W_{2}+W_{4}$ facilitates metastable surface
orientations, analogs of the strongly supercooled states. Their
transformations into the stable states are hindered by the barriers featured
by $f_{P}$ and $f_{H}$, and by the surface defects of line tension $\sim K$
that separate the areas with a different director tilt. These defects are
seen as cusps in $R\left( x\right) $, Fig.\ref{Retardation_map}a. The nuclei
of new alignment should be of a size $\sim K/\left\vert \Delta f\right\vert $
or larger, to overcome the nucleation barrier $\sim K^{2}/\left\vert \Delta
f\right\vert $. The maximum $\Delta f$ is$\ W_{2}+W_{4}$. With $W_{2}+W_{4}$
$\sim 10^{-6}$ J/m$^{2}$, the nucleation barrier is large, $\sim
K^{2}/\left( W_{2}+W_{4}\right) \sim 10^{-16}$ J $>>k_{B}T$, which signals
that the metastable states can be long-lived and that nucleation is
heterogeneous, assisted by inhomogeneities, Fig.\ref{Textural_evolution}.

The general expression $R={\int\limits_{0}^{d}{\left\vert n_{o}-{n_{eff}}%
\right\vert dz}}$ matches the data in Fig.\ref{Phase_retardation_15} at $%
d<d_{c}$, with $\alpha $=0 in the H state and $\alpha \approx $ $\Delta n_{B}%
{=0.015}$ in the P case. For the deformed states at $d>d_{c}$, the standard
model is that $R\left( d\right) $ is determined by smooth variations of $%
\widehat{\mathbf{n}}$ with $S=const$ \cite{Barbero}. Then for the A case, $%
R_{A}\left( \theta _{P}\right) =$ $dJ(\theta _{P},{0})/I(\theta _{P},{0})$,
while for the B case, $R_{B}\left( \theta _{H}\right) =dJ(\pi /2,\theta
_{H})/I(\pi /2,\theta _{H})$, where $J\left( \zeta ,{\eta }\right)
=\int\limits_{\zeta }^{{\eta }}\sqrt{K\left( \theta \right) }\left( n_{o}-{%
n_{eff}}\right) d\theta $, $I\left( \zeta ,{\eta }\right)
=\int\limits_{\zeta }^{{\eta }}\sqrt{K\left( \theta \right) }d\theta $ \cite%
{Barbero}. For ${K_{1}}={K_{3}=K}$, using the smallness of $\Delta n/n_{0}$,
one finds $R_{A}\left( d\right) =$ $d\Delta n_{A}\left( \theta _{P}-\sin
2\theta _{P}\right) /2\theta _{P}$ and $R_{B}\left( d\right) =$ $d\Delta
n_{B}\left( \pi -2\theta _{H}+\sin 2\theta _{H}\right) /\left( 2\pi -4\theta
_{H}\right) $. \ Using the discontinuities of $R$ in Fig.\ref%
{Phase_retardation_15}, $($97$\pm 3)$ nm for A and $($64$\pm 2)$ nm for B,
we conclude that the jumps in $\theta _{P}$ and $\theta _{H}$ at $d_{c\text{
}}$are substantial, $\approx ($1.3-$1.57)$, as expected from the model
above. The tilts $\alpha $ are much harder to describe as these are affected
by the tilt of boundaries between different states. Assuming a vertical
boundary for the cell B and the expression for $R_{B}\left( d\right) $
above, one finds that at $d>d_{c}$, $\alpha _{B}=\partial R_{B}/\partial d=$
$\frac{\Delta n_{B}}{2}\left( 1+\frac{\pi ^{2}}{24}\frac{K^{3}}{%
W_{2H}^{3}d^{3}}\right) =0.0075$ or larger. The experimental $\alpha _{B}=$ $%
0.0067$ is by 10\% smaller, Fig.\ref{Phase_retardation_15}. The difference
can be accounted for by the fact that ${K_{1}}<{K_{3}}$. Numerical
evaluation of $R$ with ${K_{1}}\neq {K_{3}}$\ shows that $\alpha _{B}=$ $%
0.0067$ corresponds to ${K_{1}}/{K_{3}\approx 0.4}$, a reasonable result
\cite{Nastishin_ELC}. The discrepancy between the theoretical $\alpha
_{A}\approx $ $\frac{\Delta n_{A}}{2}\left( 1-\frac{K^{2}}{2W_{2P}^{2}d^{2}}%
\right) \lessapprox $ 0.01 and the experimental $\alpha _{A}\approx $ 0.017
is more significant. It cannot be explained by ${K_{1}}/{K_{3}\neq 1}$. To
show this, we assumed $\theta _{P}=\pi /2$ (to maximize the theoretical $%
\alpha _{A}$) and then evaluated $R\ $for ${K_{1}}/{K_{3}\neq }$ 1. By
changing the ratio ${K_{1}}/{K_{3}}$ in the range $10^{-5}$ to $10^{5}$, we
find $\alpha _{A}$ changing from 0.007 to 0.013, still smaller than the
experimental value. In principle, $\alpha _{A}\approx $ 0.017 can be
obtained by allowing a nonzero $\theta _{H}\approx 0.1$. However, with $%
\theta _{H}\approx 0.1$, the cell retardation should be much higher than in
the experiment, e.g., one should measure $R$ $\approx $140 nm at $d=8.7$ $%
\mu $m, and the actual result is only $R$ $\approx $100 nm, Fig.\ref%
{Phase_retardation_15}. We thus conclude that $R\left( d\right) $ at $%
d>d_{c} $ is affected by the tilted sharp boundaries such as the ones in
Fig.2b and Fig.\ref{FCPM}c,d. \ The boundaries with sharply varying $%
\widehat{\mathbf{n}}$ \ imply a changing $S$. The latter is expected to
happen through the "interchanging eigenvalues" of the tensor order parameter
in thin hybrid aligned cells \cite{Palffy} with $d\approx \xi _{bx}$, where $%
\xi _{bx}$ is the biaxial correlation length. In thermotropic nematics, $\xi
_{bx}=(10-100) $ nm \cite{Palffy}, but in LCLCs, $\xi _{bx}$ can be larger,
since the gradients of $S$ can be accommodated by redistribution of short
and long aggregates in the intrinsically polydisperse LCLC. The standard
approach to calculate $R\ $and $\alpha \ $is not applicable to the case with
a changing $S$.

To conclude, we demonstrated that in the LCLC, surface alignment can be both
planar and homeotropic with a strongly discontinuous transition between the
two that can be described by a double-well anchoring potential. \ Much more
work is needed to gain an understanding of LCLC behavior at the surfaces and
in the bulk, as the study suggests that the LCLC structure is affected by
kinetics of self-assembly and that the director distortions can be
accompanied by gradients of the scalar order parameter.

We are grateful to the anonymous referees for useful suggestions. The work
was supported by NSF Materials World Network on Lyotropic Chromonic Liquid
Crystals DMR076290, ARRA DMR 0906751, Ohio Research Cluster on Surfaces in
Advanced Materials, NAS of Ukraine Grant \#1.4.1B/109, Fundamental Research
State Fund Project UU24/018, and by the Ministry of Education and Science of
Ukraine, Project 0109U001062.

\end{document}